\documentclass[traditabstract]{aa}  
\usepackage{txfonts}
\usepackage{epstopdf}

\usepackage{graphicx,longtable,lscape,natbib,amssymb,amssymb,amsmath}
\bibpunct{(}{)}{;}{a}{}{,} 
\newcommand{\ltsima} {$\; \buildrel < \over \sim \;$}  
\newcommand{\gtsima} {$\; \buildrel > \over \sim \;$}  
\newcommand{\lta} {\lower.5ex\hbox{\ltsima}}  
\newcommand{\gta} {\lower.5ex\hbox{\gtsima}}  
\newcommand{\ha} {H$\alpha$}  
\newcommand{\hb} {H$\beta$}

\newcommand{\kms}{$\rm{\,km \,s}^{-1}$}

\newcommand{\forb}[2]{\mbox{$[{\rm #1\, #2}]$}}
\newcommand{\oiii}{\forb{O}{III}}
\newcommand{\oi}{\forb{O}{I}\,}
\newcommand{\sii}{\forb{S}{II}\,}
\newcommand{\nii}{\forb{N}{II}\,}

\usepackage{color}

\begin{document}

\title{The VLT/MUSE view of the central galaxy in Abell 2052}
   \subtitle{Ionized gas
  swept by the expanding radio source} \author{Barbara
  Balmaverde\inst{1,2} \and Alessandro Capetti\inst{2} \and Alessandro
  Marconi\inst{3,4} \and Giacomo Venturi\inst{3,4}}

\institute {
  Scuola Normale Superiore, Piazza dei Cavalieri 7, I-56126 Pisa, Italy
  \and INAF - Osservatorio Astrofisico di Torino, Via Osservatorio 20,
  I-10025 Pino Torinese, Italy
  \and Dipartimento di Fisica e Astronomia, Universit\`a di Firenze, via
  G. Sansone 1, 50019 Sesto Fiorentino (Firenze), Italy
  \and INAF - Osservatorio Astrofisico di Arcetri, Largo Enrico Fermi 5, I-50125 Firenze,
  Italy
  }
\offprints{balmaverde@oato.inaf.it} 

\date{} 

\abstract{ We report observations of the radio galaxy 3C317 (at z=0.0345)
  located at the center of the Abell cluster A2052, obtained with the VLT/MUSE
  integral field spectrograph. The Chandra images of this cluster show
  cavities in the X-ray emitting gas, which were produced by the expansion of the radio
  lobes inflated by the active galactic nucleus (AGN). Our exquisite MUSE data
  show with unprecedented detail the complex network of line emitting
  filaments enshrouding the northern X-ray cavity. We do not detect any
  emission lines from the southern cavity,  with a luminosity asymmetry between
  the two regions higher than $\sim$ 75. The emission lines produced by the
  warm phase of the interstellar medium (WIM) enable us to obtain unique
  information on the properties of the emitting gas. We find dense gas (up to
  270 cm$^{-3}$) that makes up part  of a global quasi spherical outflow  that is driven by
  the radio source, and obtain a direct estimate of the expansion velocity of
  the cavities (265 km s$^{-1}$). The emission lines diagnostic rules out
  ionization from the AGN or from star-forming regions, suggesting instead
  ionization from slow shocks or from cosmic rays.  The striking asymmetric
    line emission observed between the two cavities contrasts with the less
    pronounced differences between the north and south sides in the hot gas;
    this represents a significant new ingredient for our understanding of
    the process of the exchange of energy between the relativistic plasma and the
    external medium.  We conclude that the expanding radio lobes displace the
  hot tenuous phase of the interstellar medium (ISM), but also impact the
  colder and denser ISM phases.  These results show the effects of the AGN on
  its host and the importance of radio mode feedback. } \keywords{galaxies:
  active -- galaxies: nuclei -- galaxies: clusters: general -- galaxies:
  clusters: individual (A2052) -- galaxies: star formation --ISM: jets and
  outflows}

\titlerunning{The VLT/MUSE view of the central cluster galaxy 3C317 } 
\authorrunning{B. Balmaverde et al.}
 \maketitle

\section{Introduction}
\label{intro}

The exchange of matter and energy between active galactic nuclei (AGN), their
host galaxies, and clusters of galaxies--known as the  AGN feedback process--is a fundamental ingredient in the formation and evolution of astrophysical
structures. Galaxy formation models consider two modes of AGN feedback: the
quasar mode, which operates during a luminous nuclear phase and produces winds
powered by radiation pressure, and the radio mode, in which kinetic energy is
released through relativistic jets \citep{fabian12}. Powerful outflows have
been extensively observed at both low and high redshift
\citep{swinbank15,carniani16,maiolino12}, but the observational evidence that
they have a relevant impact on the star formation of their host galaxies
remains elusive. 

So far, the clearest evidence of AGN feedback has been found in the local
Universe, where X-ray images of galaxy clusters revealed cavities in the hot
 ionized medium (HIM) ($10^6 - 10^7$ K), often filled by the radio emitting
plasma. The radio sources hosted in the brightest cluster galaxies (BCGs) were
able to displace the low-density  gas forming the hot phase of the ISM while
expanding \citep{mcnamara00,birzan04,birzan12}.

The X-ray image of the galaxy cluster A2052 from the Chandra satellite provides
one of the clearest examples of such cavities: a pair of symmetric depressions
in the X-ray surface brightness is closely correlated with the presence of
radio emission \citep{blanton01}. In A2052 the estimate of the total
energy required to inflate the cavity is a few 10$^{57}$ erg
\citep{birzan04}. Since gas speed cannot be measured from X-ray observations, it
is assumed that the bubble moves outward at the sound speed ($\sim500$
\kms\ for an external gas temperature of 10$^7$ K, \citealt{blanton11}) or,
alternatively, that its motion is driven by buoyancy forces
\citep{panagoulia14}, resulting in a cavity age of $\sim1-2\times10^7$ years.

The presence of warm ionized medium (WIM) producing optical emission lines in
the central region of A2052 \citep{baum88,heckman89,mcdonald10} offers the unique
possibility of studying the dynamics of the expanding cavity and the energy
exchange between the active nucleus and the surrounding medium, an essential
ingredient for a better understanding of the feedback process.

We observed the central brightest cluster galaxy, UGC~9799, in the
optical band with the VLT/MUSE integral field spectrograph. UGC~9799 is an
elliptical galaxy (z=0.0345 where 1$"$ corresponds to 0.69 kpc) associated with
the radio source 3C~317. The mass of its central supermassive black hole,
estimated from the central stellar velocity dispersion ($\sigma_{\rm
  star}=190$ km s$^{-1}$, \citealt{smith04}), and the relation of $\sigma_{\rm
  star}$ with $M_{\rm BH}$ \citep{kormendy13} is $M_{\rm BH}=$10$^8
M_\odot$. Source 3C~317 shows an amorphous radio halo, elongated in the NS direction
and extended over $\sim$60 kpc \citep{morganti93}, with a radio luminosity of
P=$1.3 ~ 10^{33} $erg s$^{-1}$ Hz$^{-1}$ at 178 MHz \citep{spinrad85}; the radio core
is coincident with the optical center of the galaxy.  The optical nucleus has
emission line ratios typical of low-ionization nuclear emission line regions
(LINERs, \citealt{heckman80,buttiglione10}).

\section{Observation and data reduction}
\label{sample}

We pointed the telescope at A2052 and we took advantage of the large field of
view ($\sim1'\times1'$) of the instrument to observe both the northern and
southern X-ray cavities.  The observations were obtained with the VLT/MUSE
spectrograph on 13 April, 13 May, and 28 May 2016 for about two hours split between
five exposures of $\sim$25 minutes each, taken with different seeing
conditions (between 0.6$\arcsec$ to 2.0$\arcsec$). Four datasets were affected by 
bad sky conditions, resulting in strong reddening and uncertainties in the
flux calibration or poor seeing conditions. We thus decided to limit our
analysis to the exposure with the highest spatial resolution (0.6$\arcsec$)
and best sky conditions, namely MUSE.2016-05-13T05:19:36.833. We used the ESO
MUSE pipeline (version 1.6.2) to obtain a fully reduced and calibrated data
cube.  No correction for the small Galactic reddening, E(B-V)=0.03, is
  applied.

To subtract the stellar continuum, we need to optimally bin the data to detect
the stellar absorption features in the spectra, preserving the highest
possible spatial resolution. Therefore, we performed a Voronoi adaptive
spatial binning of the cube data, requiring an average signal-to-noise ratio
per wavelength channel of at least 50 in the range 4500-7000 $\AA$. We used
the Penalized Pixel-Fitting code (pPXF, \citealt{cappellari03}) to fit the
absorption stellar features with a linearly independent subset of stellar
templates from \citealt{vazdekis10} that combine theoretical isochrones with
the full MILES empirical stellar library (\citealt{falcon11}). We adopt a
10th order additive polynomial correction for the continuum and included  the emission lines in
the fit instead of masking them.  From this analysis
  we also estimated the recession velocity for UGC~9799 of 10330$\pm$10 \kms,
  in good agreement with previous optical measurements
  \citep{katgert98,wegner99,smith00}.

We fit simultaneously all emission lines in the red portion of the spectrum
(from 6000 to 7000 $\AA$) with a single Gaussian profile, one for each
line. The line wavelength separation is fixed to the theoretical value, while
the intensities are free to vary (except for the \nii
$\lambda\lambda$6584,6548 ratio, which is fixed to 3). We repeated the reduction
procedure for the blue spectrum to measure the H$\beta$ and \oiii
$\lambda\lambda$4959,5007 lines. In a few regions, the emission lines show
two well-separated peaks and, in these cases, we included a second Gaussian
component.

\begin{figure}
\centering{
\includegraphics[scale=0.5,angle=0]{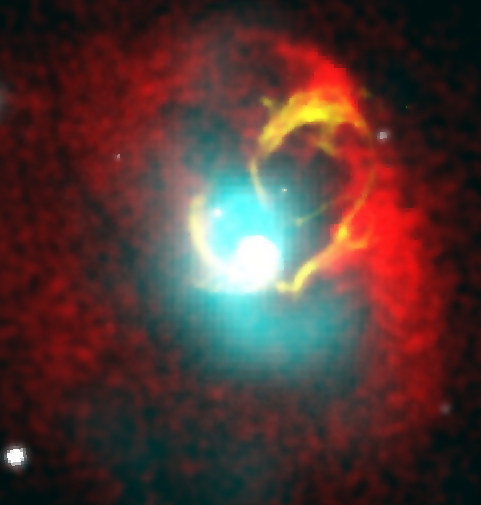}
\caption{Multiwavelength composite image of the central regions of the A2052
  cluster.  Optical continuum image in the r band from the the Sloan Digital
  Sky Survey (white), radio map from the Very Large Array (cyan),
  X-ray emission from Chandra (red) and the \ha+\nii emitting filaments
  we observed with VLT/MUSE (yellow). The image covers the MUSE field of
  view, i.e., 60$\arcsec \times 60 \arcsec$, 41.4 $\times$ 41.4 kpc.}
\label{comp}}
\end{figure}

\begin{figure}
\centering{
\includegraphics[angle=0,scale=0.430]{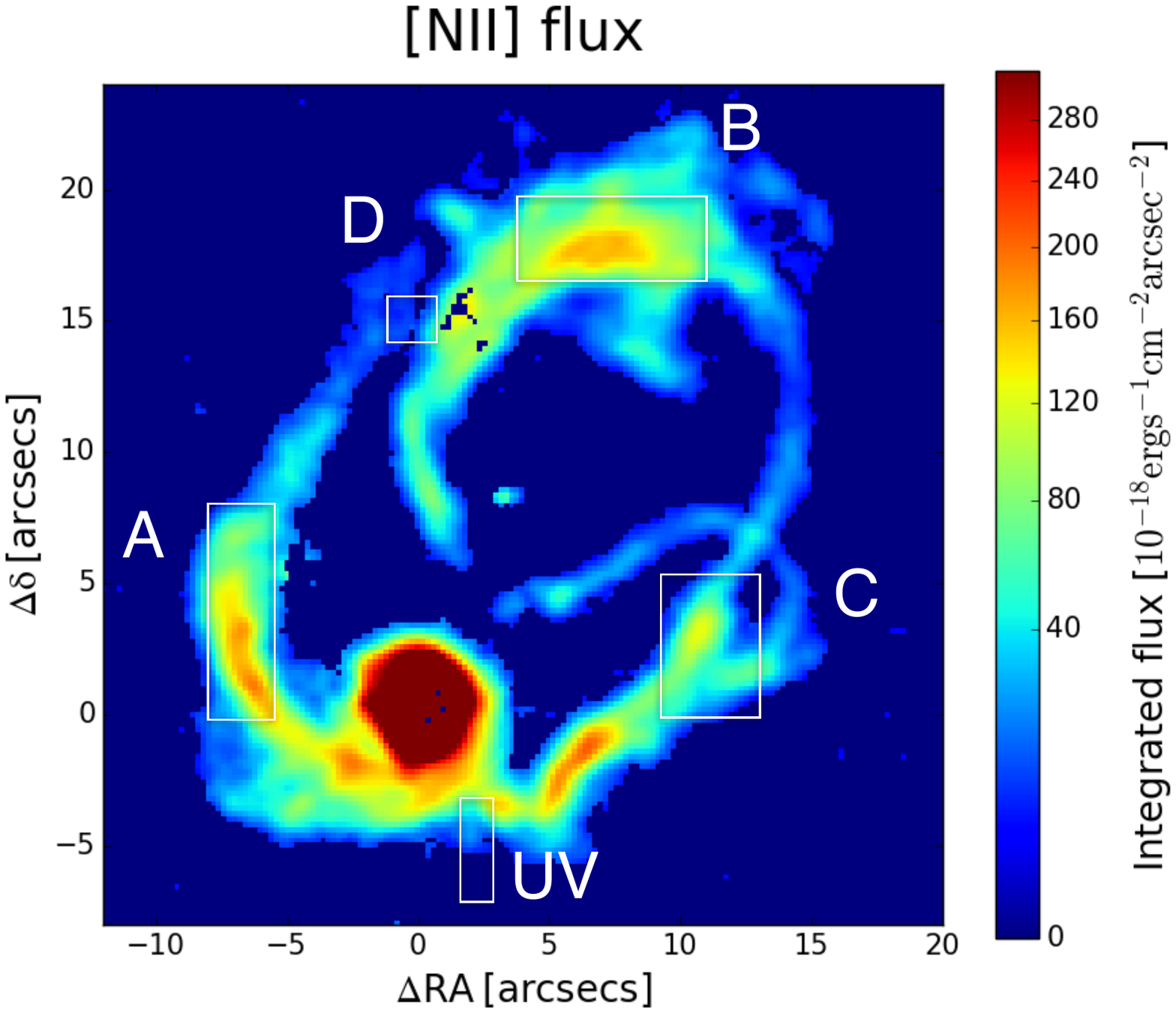}
\caption{Surface brightness of the \nii$\lambda$6584 emission line. 
The image covers the 33$\arcsec \times 33 \arcsec$, 22.8 $\times$ 22.8 kpc
}
\label{maps}}
\end{figure}

\begin{figure*}
\centering{
\includegraphics[scale=0.35,angle=0]{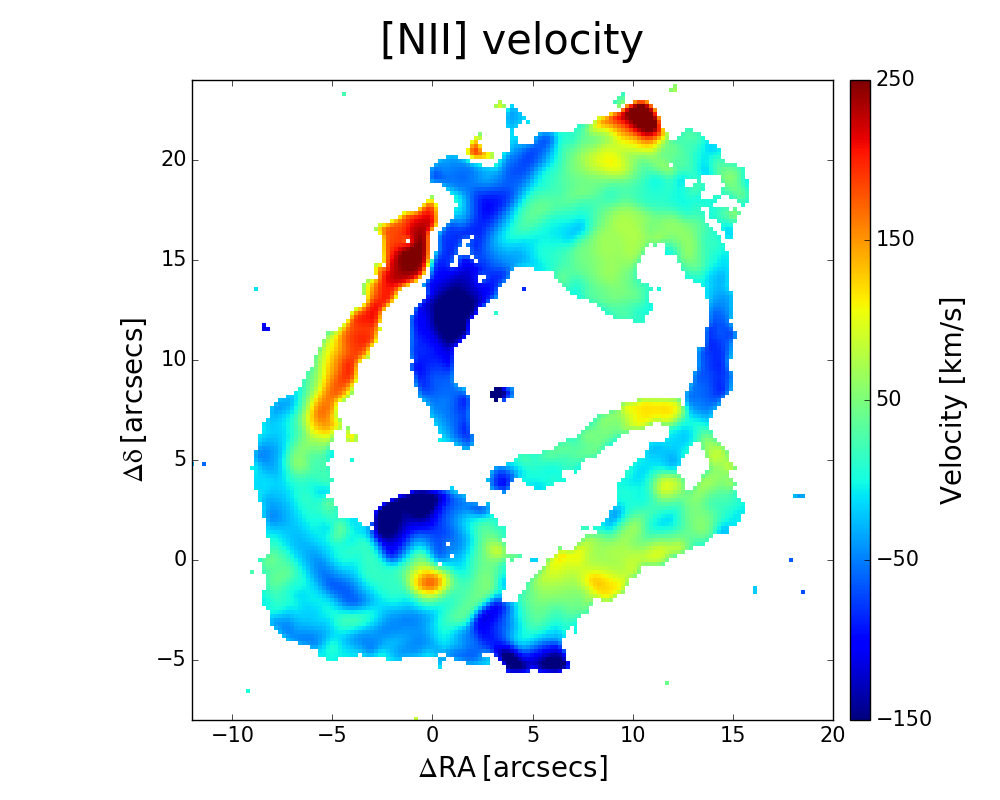}
\includegraphics[scale=0.35,angle=0]{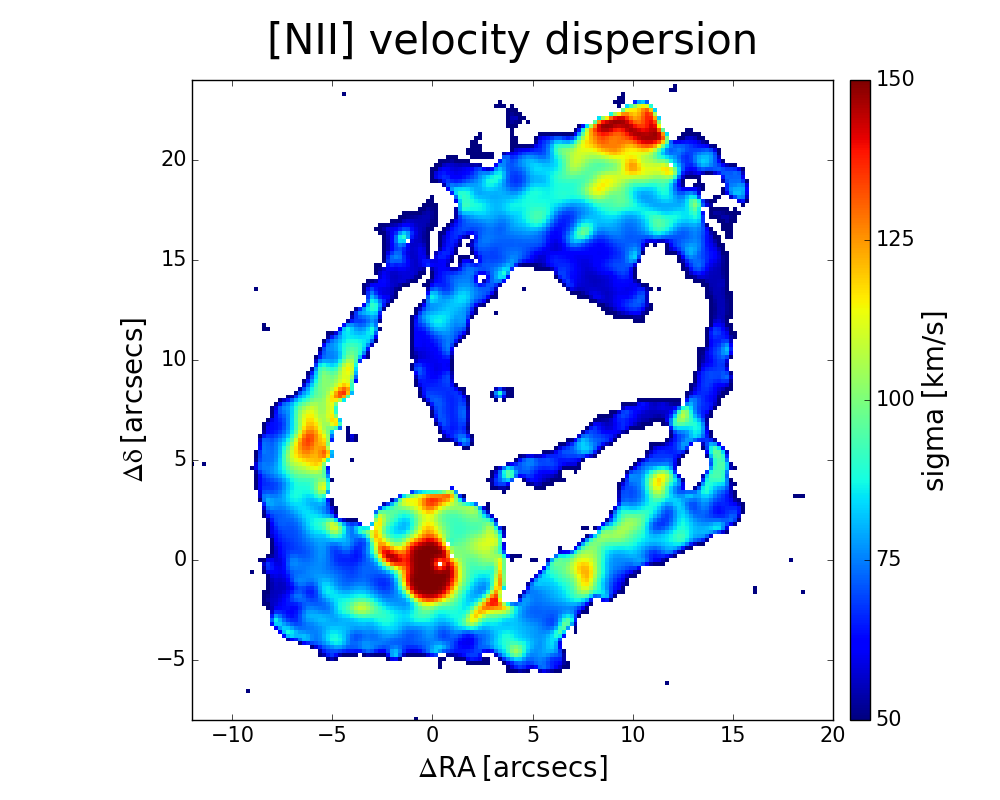}
\caption{Gas velocity (left panel) and velocity dispersion (right panel)
  derived from the \nii\ line. }
\label{niiveldisp}}
\end{figure*}

\begin{figure}
\centering{
\includegraphics[scale=0.35,angle=270]{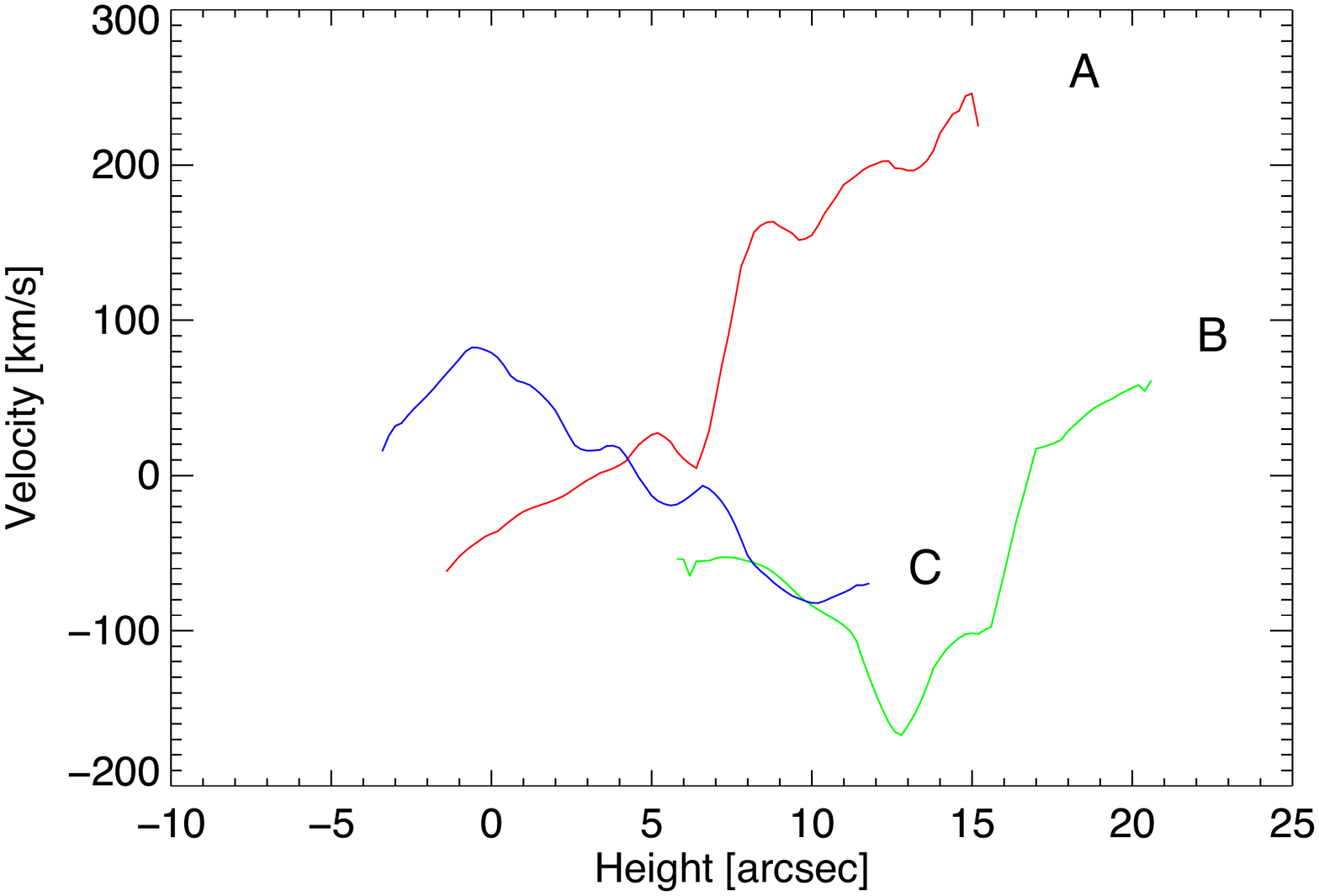}
\caption{Position--velocity diagram for the three main emission line
  filaments. At each given height, we measured the velocity at the brightness
  peak, i.e., following the filament's ridge line.}
\label{pv}}
\end{figure}

\begin{figure*}
\centering{
\includegraphics[scale=0.25,angle=0]{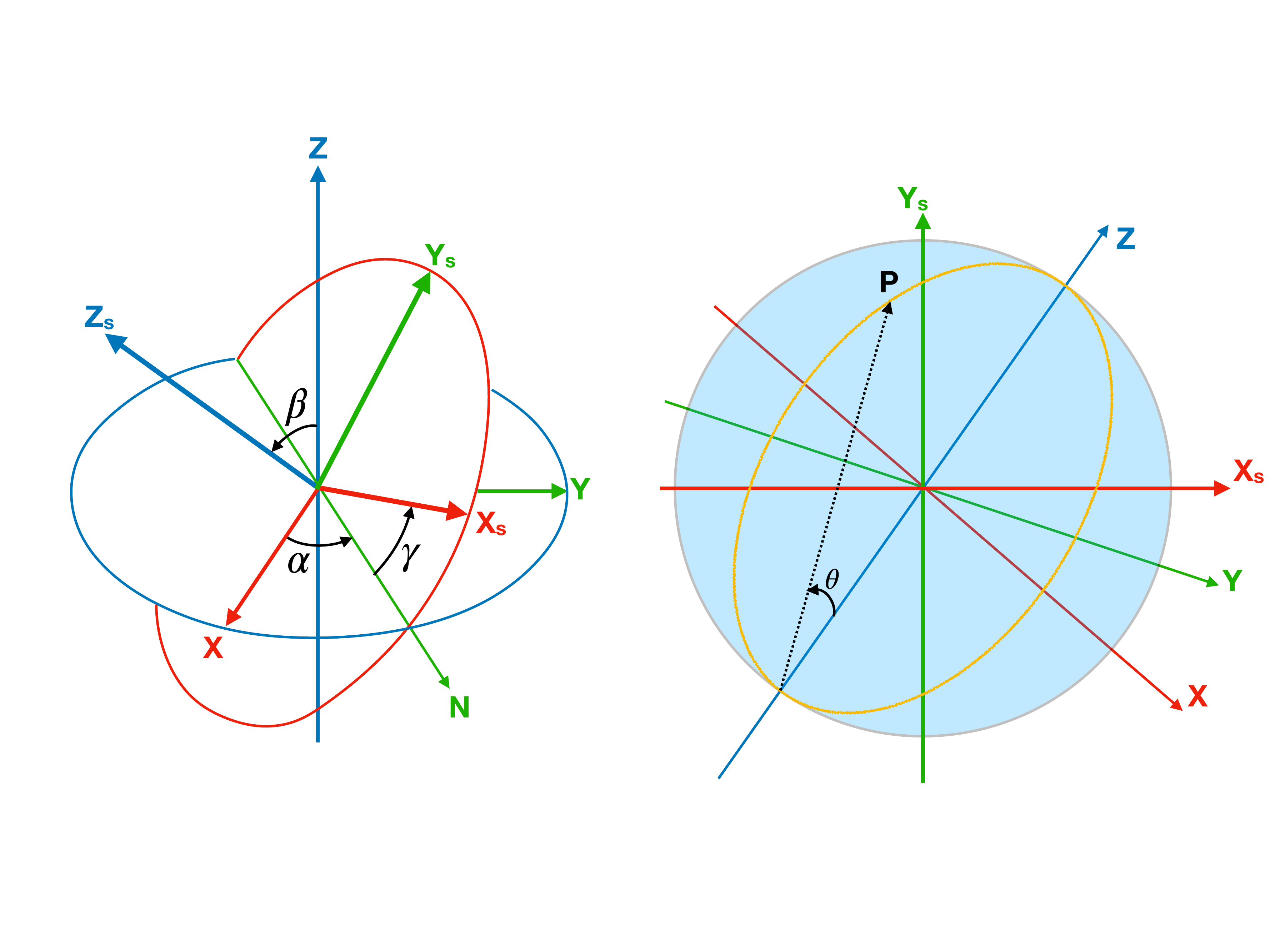}
\includegraphics[scale=0.30,angle=0]{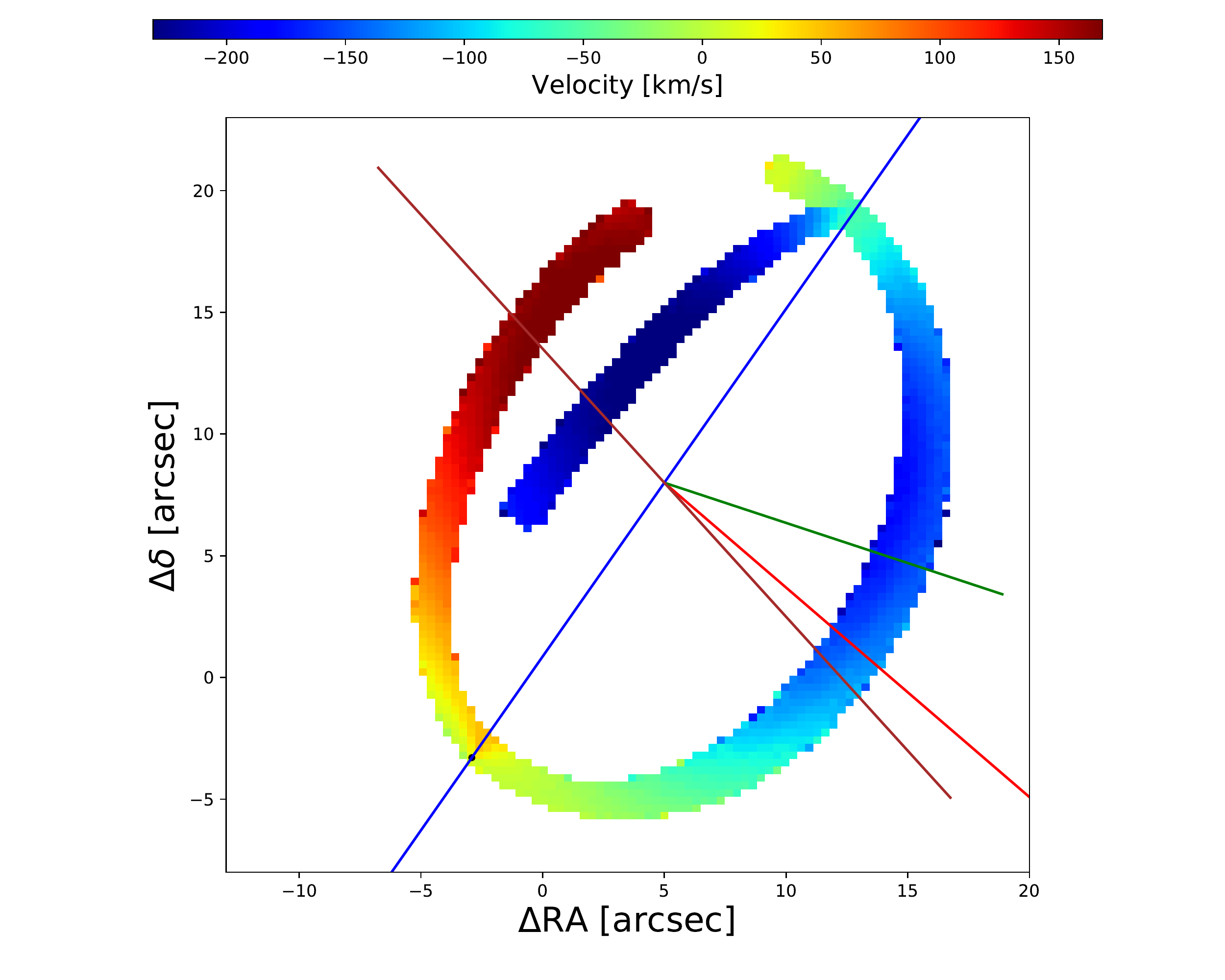}
\caption{Left and middle panel: Reference coordinate systems and rotation angles used in the kinematical model. 
Right panel: Toy model of a spherical expanding bubble. The origin of the reference coordinates is the center of the cavity. The
  velocities of the receding and approaching halves are reproduced in red and blue. The black dot marks
  the location of the bubble origin. The three reference axes in each figure are the X-axis (in red), the Y-axis (in green), and Z-axis (in blue). We plot  the ring axis in the  XY-plane (in brown). }
\label{toy}}
\end{figure*}

\begin{figure}
\centering{
\includegraphics[scale=0.35,angle=0]{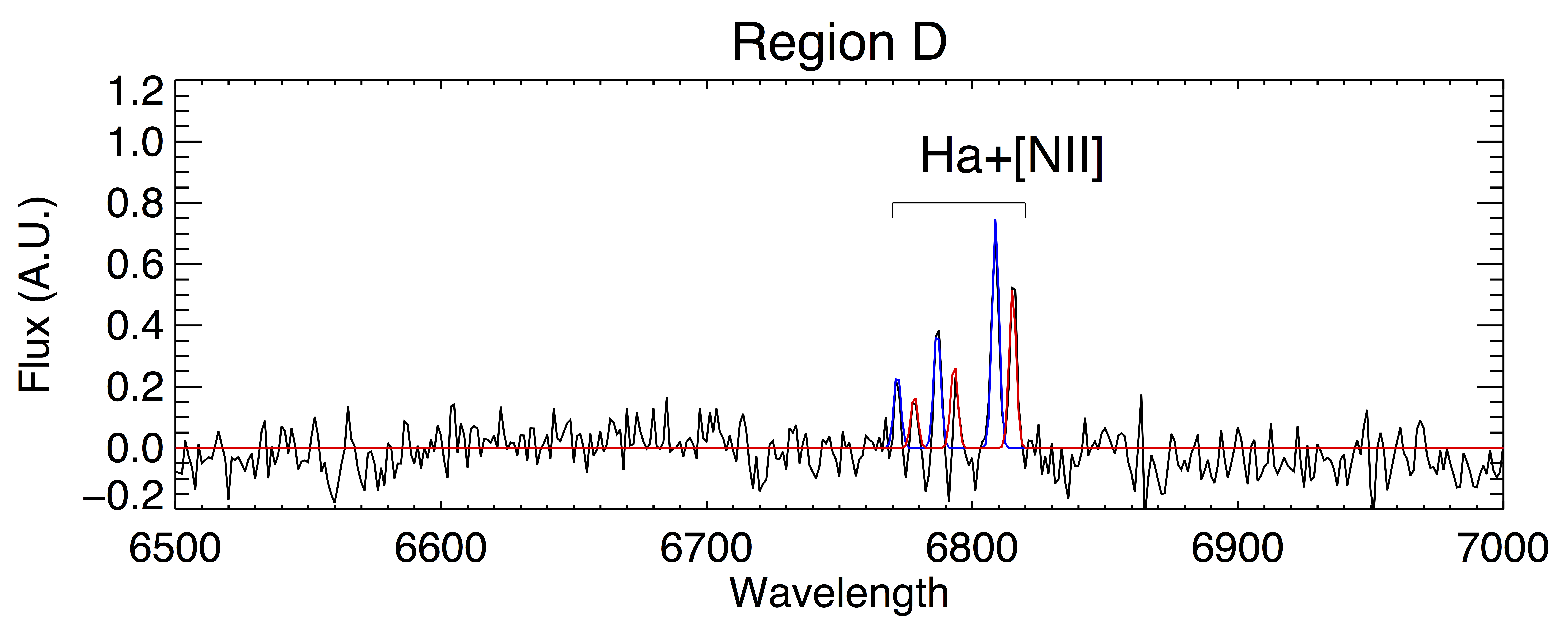}
\caption{Red spectra of region D showing one of the locations of split
  emission lines where we performed a fit with two Gaussians, represented by
  the blue and red curves.}
\label{split}}
\end{figure}

\section{Results}

The resulting image obtained in the forbidden \nii line at 6584 \AA, the
strongest line in our spectra, shows a complex network of filaments
enshrouding the northern X-ray cavity (Figs. \ref{comp} and \ref{maps}). The
emission line filaments are co-spatial with those traced by the hot gas and by
the dust features seen in the Hubble Space Telescope images \citep{sparks00},
while they are anti-correlated with the radio emitting plasma. No lines are
detected from the southern cavity. We integrated over two $20\arcsec \times
20\arcsec$ square regions centered at 13$\arcsec$ NW and SE from the nucleus,
and found a flux asymmetry higher than 75.

The velocity field of the warm gas mapped by the \nii emission
(Fig. \ref{niiveldisp}, left panel) is rather complex, but the hallmarks of the
cavity's expansion are clearly visible. Starting immediately south of the
central galaxy, a bright filament (A in  Fig. \ref{maps}) shows a velocity
gradient starting from a null velocity and reaching a redshift of $265\pm8$
\kms\ (see Fig. \ref{pv}). This is likely due to the gradual increase in
the projected velocity along the line of sight as the filament wraps around
the cavity. Filament A intersects another emission line structure (B in
Fig. \ref{maps}) that shows a similar but opposite behavior, reaching a
maximum blueshift of $-180\pm2$ \kms. The emission from these two filaments (A
behind the cavity, B in front of it) is projected onto the same location in
the sky producing a region of split-lines (D in Fig. \ref{maps}), with a
velocity separation of $290\pm10$ \kms\ (Fig. \ref{split}), an effect that also
occurs  in other regions of the nebula. This is the characteristic
signature of an expanding bubble when both sides are simultaneously visible.

 The overall gas kinematics are well reproduced by the toy model of an
  expanding bubble presented in Fig. \ref{toy}. The bubble is surrounded by line emitting filaments modeled  as parts of thick polar rings with the same radius but different orientations. The model confirms that, in general, we see only one of the two velocity components of an expanding filament (partial ring).
  In a cartesian coordinate system (x, y, z) we consider spherical coordinates (r, $\theta$, $\phi$). 
  We simulate each filament considering a ring with thickness r, r+dr with null velocity in the galactic nucleus (in x=0, y=0, z=-r). We assume that in each filament the intrinsic
  expansion velocity is higher toward the pole and decreases at lower
  latitudes to produce the quasi-spherical shape $v=v_0*\cos\theta$, with $\theta$ being the angle between the $z$-axis  and the direction from the galaxy nucleus (x=0, y=0, z=-r) to the point on the sphere. 
 To project onto the plane of the sky (X,Y) taking into account  the orientation with respect to the line of sight, we consider a rotation of the system around 
 the three Euler angles $\alpha$, $\beta$, $\gamma$ obtaining a new system of reference  (X,Y, Z) in which the Z-axis  is oriented along the line of sight. We locate the origin of the reference frame 
 at the center of the cavity.
We obtain a good match to the observation assuming  $\alpha$=30$^\circ$, $\beta$=70$^\circ$, $\gamma$=35$^\circ$ with 
different $\phi$ ranges for all  three filaments. 
The velocities along the
  line of sight of the receding and approaching halves are shown in red and blue. Several of the observed features are accounted for: e.g,
  the increase in velocity along filament A, the highly negative velocity at
  the pole, the velocity structure of filament B (reaching the maximum
  speed at the bubble's center), and the general
  shallower gradients along the west with respect to the east side.

The line widths (Fig. \ref{niiveldisp},
right panel) are generally between 50 and 100 \kms, consistent with MUSE
instrumental resolution; higher values are found, besides the central $\sim$1
kpc of the host, only over rather small regions and notably at the cavity's NW
tip where line widths up to 150\kms are reached.

The emission line intensity ratios can be used to distinguish between
different ionization mechanism of the nebular gas: H~II regions, AGN, or
shocks \citep{baldwin81,dopita95}. To explore the physical conditions of the
WIM, we modeled the other weaker emission lines present in the spectra,
i.e., \hb, \ha, \oiii$\lambda$5007, \oi$\lambda$6300, and the \sii doublet at
$\lambda\lambda$6716,6731.

\begin{figure*}
\centering{
\includegraphics[scale=0.35,angle=0]{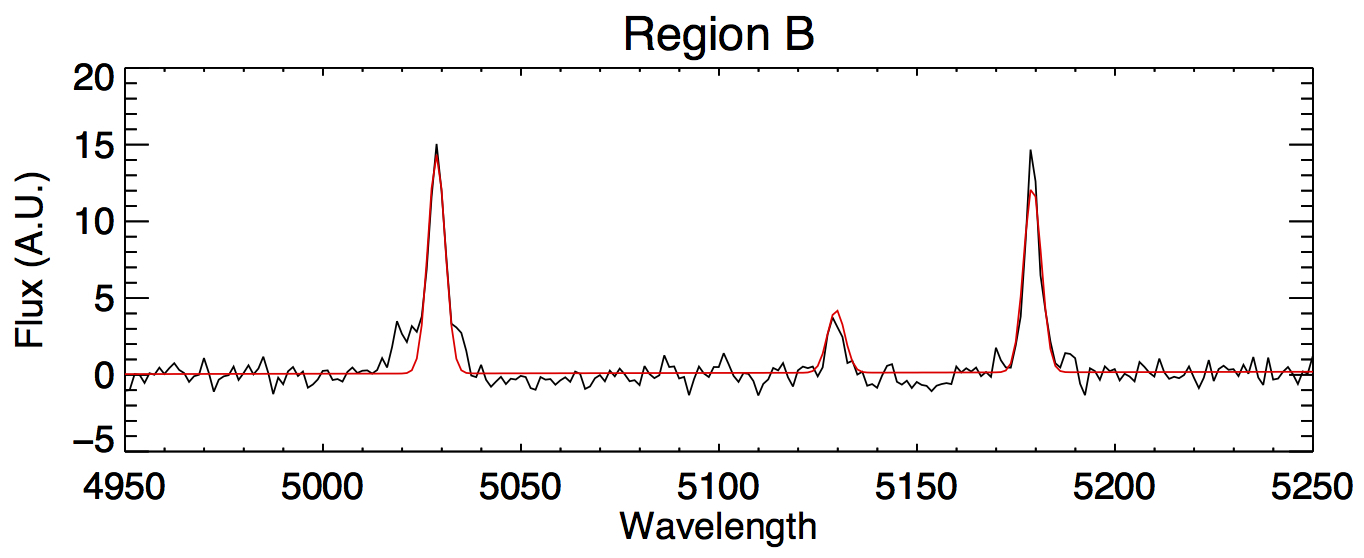}
\includegraphics[scale=0.35,angle=0]{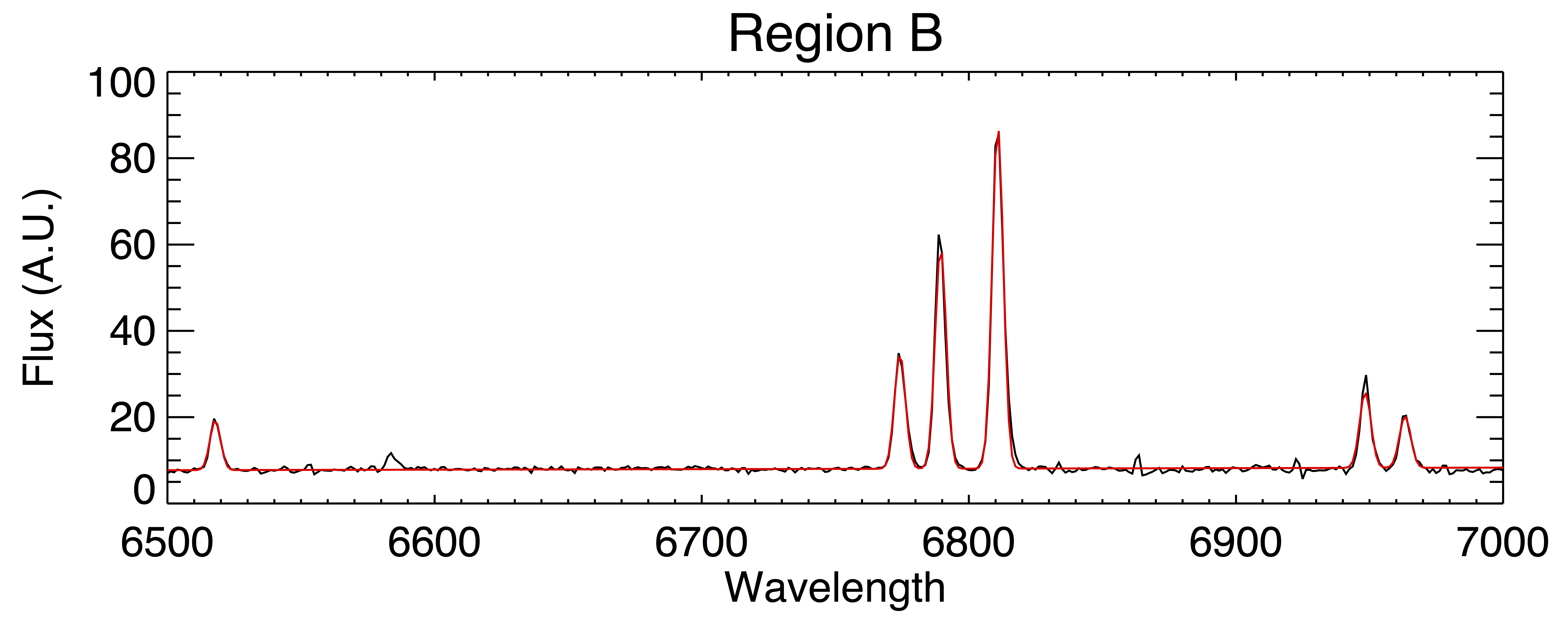}

\includegraphics[scale=0.35,angle=0]{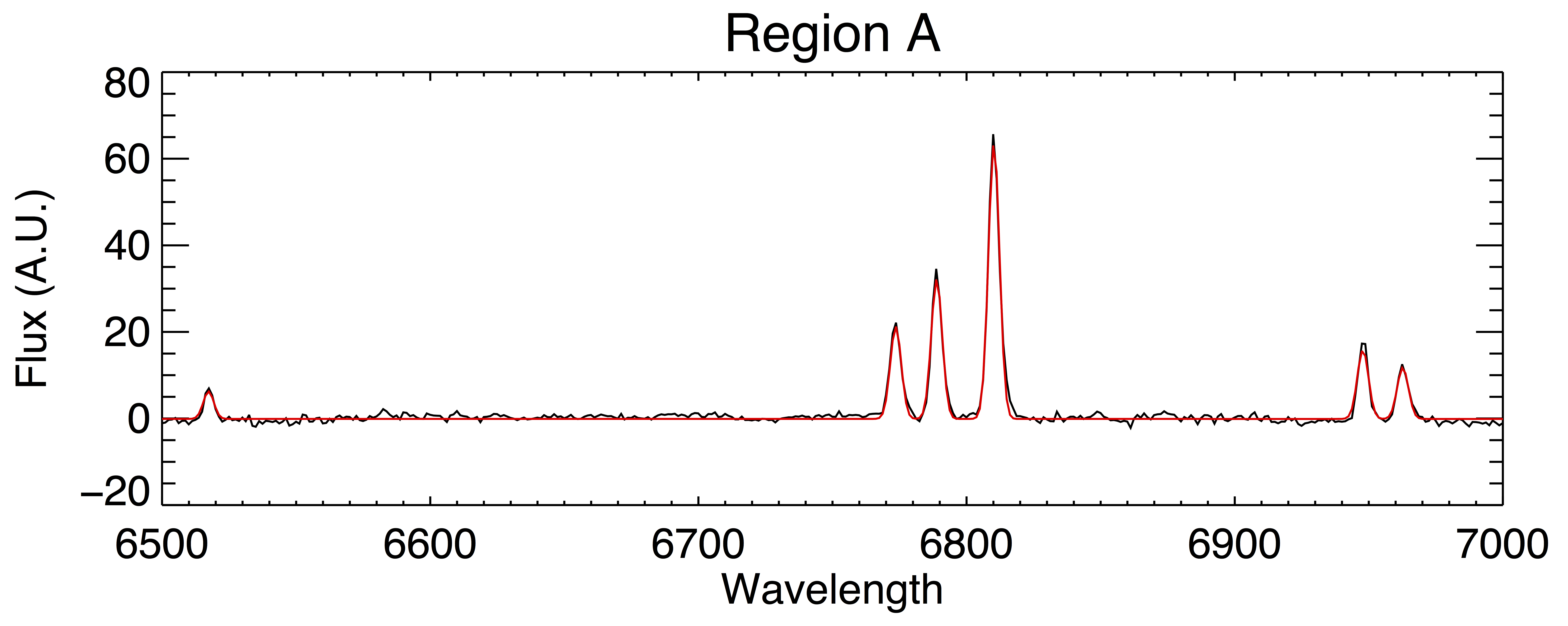}
\includegraphics[scale=0.35,angle=0]{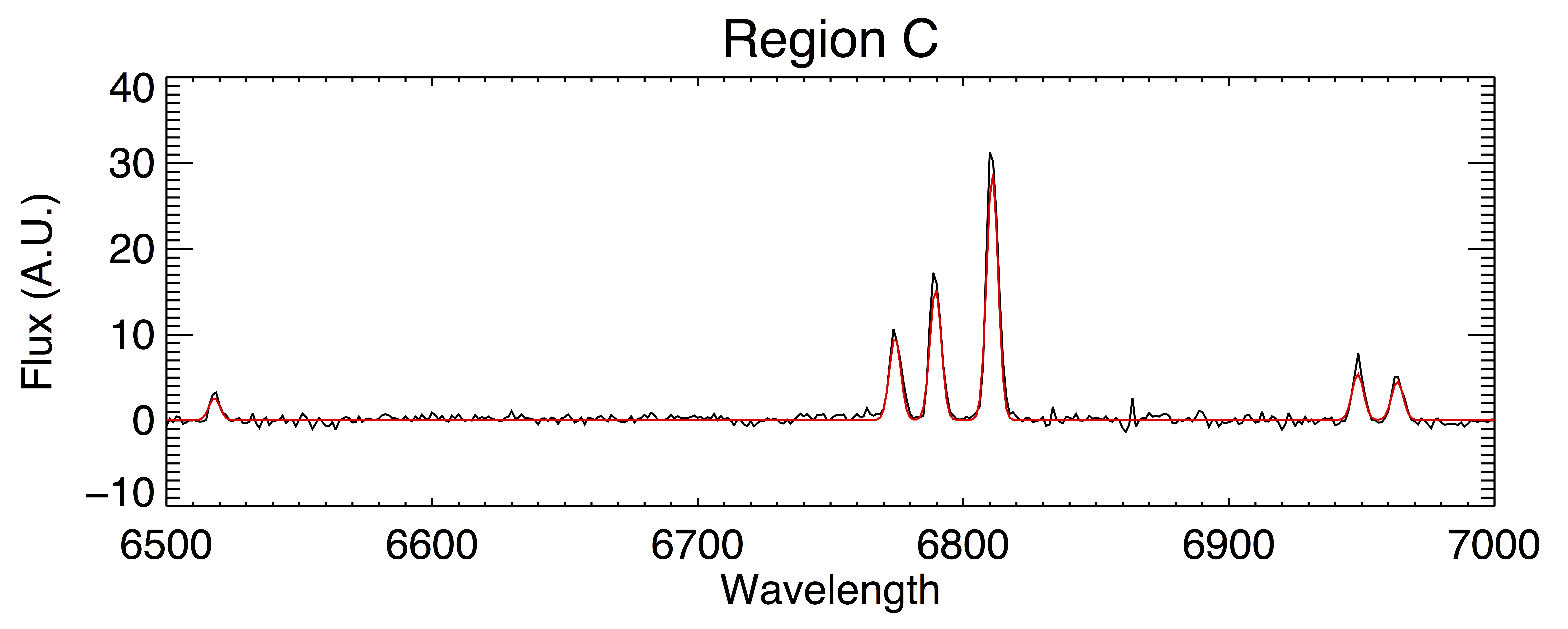}

\caption{Top panels: Blue and red spectra of region B. Bottom 
  panels: Red spectra of regions A and C. 
}
\label{spettri}}
\end{figure*}

\begin{figure*}
\centering{
\includegraphics[width=18.0cm,angle=180]{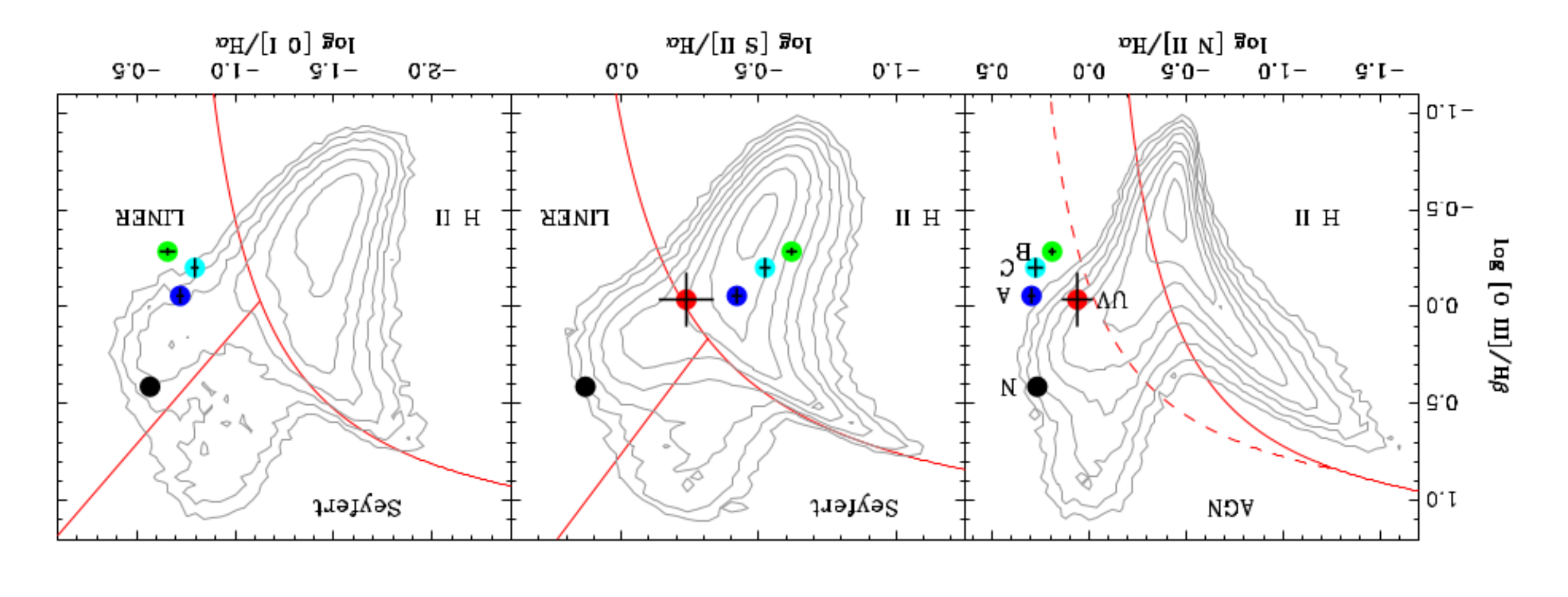}
\caption{Location of regions A, B, C, nucleus (N), and UV filament (green,
  cyan, blue, black, and red dots, respectively) in the spectroscopic
  diagnostic diagrams.  The solid lines separate star-forming galaxies,
  LINERs, and Seyferts \citep{kewley06}. Contours represent the iso-densities
  of all SDSS/DR7 emission line galaxies \citep{capetti11}.  }
\label{ratios}}
\end{figure*}

In Fig. \ref{spettri} we show the spectra of regions A, B, and C. In the plane
defined by the \nii/\ha\ and \oiii/\hb\ ratio (Fig. \ref{ratios}), their
representative points fall into a region not populated by any of the emission
line galaxies extracted from the SDSS DR7 data \citep{kewley06,capetti11}: in
these objects the dominant ionization mechanism is due to the high-energy
photon field produced by young stars or by the active nucleus. The
inconsistency with the properties of photoionized gas in 3C~317 is confirmed
by the other two diagrams where a contradictory classification would be
obtained, as star-forming  H~II regions from the \sii/\ha\ ratio and as LINER
from \oi/\ha. This suggests that the gas is ionized by a different mechanism.

\begin{figure*}
\centering{
\includegraphics[scale=0.35,angle=0]{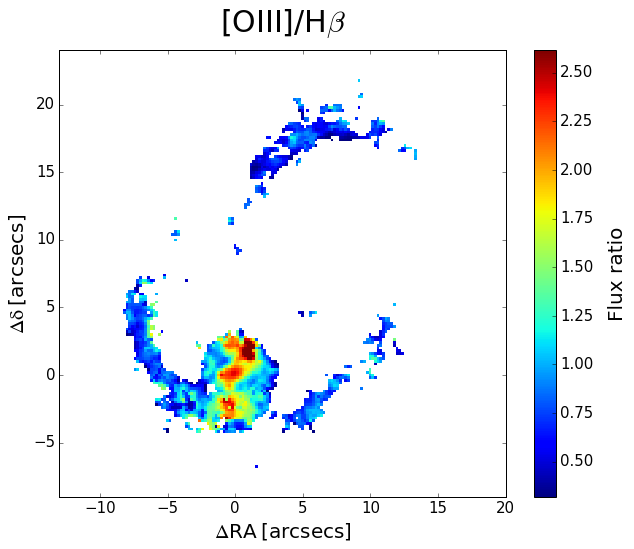}
\includegraphics[scale=0.35,angle=0]{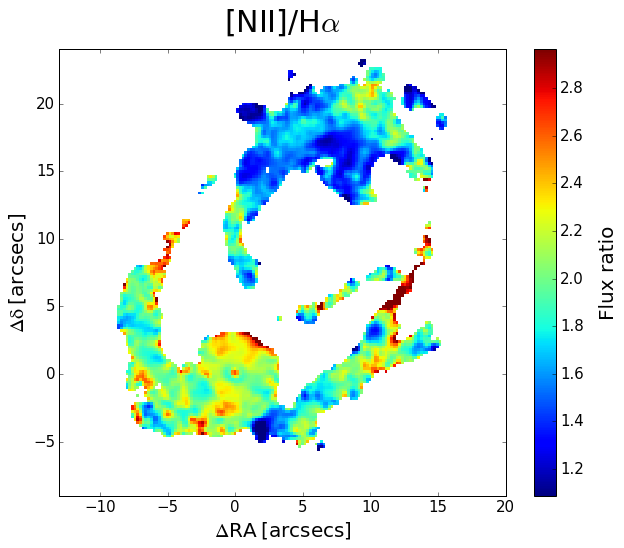}
\includegraphics[scale=0.35,angle=0]{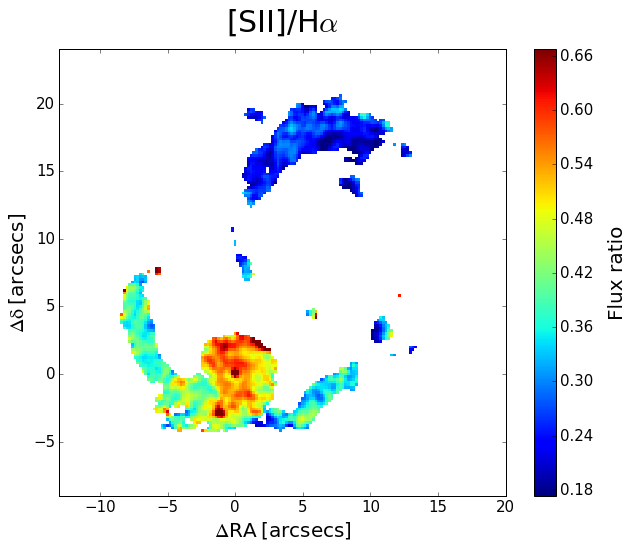}
\includegraphics[scale=0.35,angle=0]{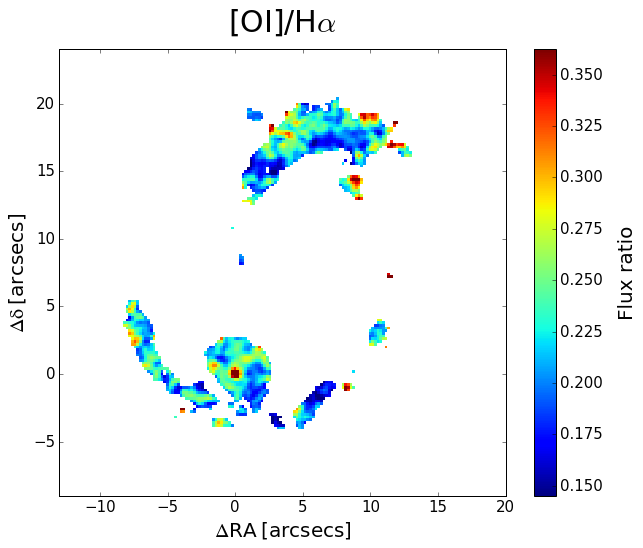}
\caption{Spatial map of the diagnostic line ratios.}
\label{niiha}}
\end{figure*}

This result is confirmed overall across the nebula because we measured rather
constant values (\oiii/\hb $\sim 0.5-0.9$, \nii/\ha $\sim1.5-2.0$, \sii/\ha
$\sim 0.2-0.4$, and \oi/\ha $\sim 0.15-0.25$), which again rules out the AGN
or young stars as the dominant sources of ionization of the gas. In particular the ratio  
\nii/\ha\  is everywhere higher than 1.2 (see Fig. \ref{niiha}),
inconsistently with the values predicted for and observed in star-forming
regions \citep{kewley01}. The only exception is a small area (located
  2$\arcsec$ S and 4 $\arcsec$ W of the host where  \nii/\ha\  is
  $\sim$ 0.8), which  we discuss in more detail in Sect. 4.

We also estimated the gas density from the ratio $R_{\sii}$ of the \sii lines
at 6716 and 6731 $\AA$ \citep{osterbrock89} in the same three regions of the
nebula considered above, obtaining $R_{\sii,A} = 1.34\pm0.05$ in region A,
$R_{\sii,B} = 1.41\pm0.04$ in region B, and $R_{\sii,C} = 1.18\pm0.05$ in
region C. By adopting a temperature of 10$^4$ K, the corresponding gas
densities are $n_{\rm e}=270^{+70}_{-70}\,{\rm cm}^{-3}$ and $n_{\rm
  e}=80^{+50}_{-45}\,{\rm cm}^{-3}$ in C and A, respectively, while only an
upper limit of $n_{\rm e}=50\,{\rm cm}^{-3}$ can be set on region B.

The total mass of the ionized gas can be estimated as
\begin{equation*}
M=7.5\times10^{-3}\left(\frac{10^4}{n_e}\frac{L_{H\beta}}{L_{\odot}}\right) M_{\odot},
\end{equation*} 
where $L_{\rm H\beta}$ is the luminosity of the \hb\ emission line
\citep{osterbrock89}.  Adopting a density of $n_{\rm e}=100\,{\rm cm}^{-3}$ we
integrated the \ha\ emission line map (excluding a circular region of
4$\arcsec$ of radius centered on the host galaxy and assuming
$\frac{F_{H\beta}}{F_{H\alpha}}=3$), finding a total mass of $M_{\rm tot} =
1.3\times10^6$ M$_\odot$.  We then estimated the total kinetic energy of the
gas as $\frac{1}{2}M\,v^2$, where $M$ is the mass estimated in each spatial bin
and $v$ is the corresponding observed velocity. Integrating all over the
nebula we derive a total kinetic energy in the WIM of $1.6\times10^{53}$ erg.
In this calculation there are two main sources of uncertainty: 1) the effects
of projection on the velocity field and 2) the assumption on the gas density
value because in some regions the \sii doublet lines ratio is close to the low-density limit value and only returns to upper limits for the density  in
region B. The reported gas mass and kinetic energy are likely to be
underestimated and should only be considered   an order of magnitude
estimate. These values indicate that the acceleration of the WIM gas only
requires the conversion of a small fraction ($\sim10^{-4}$) of the total
energy injected into the ISM by the AGN ($\sim10^{57}$ erg,
\citealt{birzan04}), thus not representing a significant demand for the
efficiency of the process of energy exchange. However, the gas in the warm
phase represents only a very small fraction of the ISM, which is by far
dominated by the hot phase and, in many brightest cluster galaxies (BCGs),
also has an important contribution of molecular gas (e.g.,
\citealt{salome03}).

\section{Discussion}

The expansion rate of the cavities inflated by the radio lobes cannot be
measured from X-ray observations: both the age and the power required are
derived from indirect and model-dependent methods. MUSE observations
instead enable us to directly measure  the expansion speed of the cavity in
A2052. The line emitting gas is located at the edges of the northern cavity
and it moves with an ordered velocity field tracing a quasi-spherical
outflow. The maximum observed speed is 265 \kms\ measured at the location of
the A filament, 10 kpc from the nucleus. This value is lower with respect to
the estimates obtained from X-ray observations, $\sim 600-700$
\kms\ \citep{birzan04}. However, since the optical nebula only partly covers 
the cavity we might still suffer from projection effects. Furthermore, the
X-ray data return an average of the expansion speed over its lifetime, while
the optical data are instantaneous measurements and the expansion of
over-pressured radio lobes slows with time \citep{begelman89}. A similar
comparison between optical and X-ray observations should be extended to other
similar sources to further explore this issue.

We found that while the emission line ratios on the nucleus are typical of a
LINER, in the filaments they are inconsistent with ionization from the AGN or
young stars. The transition from the nuclear LINER spectrum to that typical of
the nebula occurs at the very onset of the filaments, at $\sim$ 3 kpc from the
nucleus. This is clearly marked by a decrease in  \oiii/\hb\  from the
central value of $\sim 3$ to $\lesssim 1$ and in  \sii/\ha\  from
$\sim 0.6$ to $\sim$0.3. At this distance the AGN ionization leaves place for a
different mechanism. There are two possible alternatives: ionization due
to slow shocks \citep{dopita95} or collisional heating from cosmic rays
\citep{ferland08,ferland09,fabian11}. The small width of the emission lines,
$\sim 50 - 100$ \kms, argues against the importance of shocks particularly
noting that the broader lines are found in regions of large velocity
gradients  and where different filaments intersect, thus favoring ionization
from energetic particles.

Various techniques provide similar indications for a modest level  of star formation ($\sim
1
M_\odot $yr$^{-1}$) in the central regions of UGC~9799
\citep{crawford99,blanton03,odea08}. In addition, a filament of UV emission
extending from 4 to 6 kpc SW of the nucleus is interpreted as a star-forming
region with a rate of $\sim 10^{-3} M_\odot $yr$^{-1}$ \citep{martel02}.
Based on our results, we can conclude that star formation does not have a
significant impact on the line ratios and on the gas physical conditions.
  Nonetheless, the lowest value of  \nii/\ha\  over the whole
  nebula, $\sim$ 0.8, is found at the location of the brightest UV knots. The
  integrated spectrum of this region is still characteristic of LINERs, but
  its representative point in the diagnostic diagrams is located close to the
  boundary of [H~II] objects, likely due to the contribution of this star-forming region.

The depth of the MUSE observations of A2052 allowed us to set stringent limits
on any line flux from the southern side of the galaxy, with an asymmetry
higher than a factor $\sim$ 75 with respect to the northern side. This result
has implications on the origin and evolution of the warm and cold phases of
its ISM because the extreme differences in line emission contrast with
  the symmetry of the radio and starlight. In X-rays the morphology is also
  rather symmetric: within the same regions on which we estimated the line
  asymmetry, the number counts in the Chandra image returns a ratio of
  1.5. The northern cavity also shows a temperature a factor  of $\sim$ 2 lower
  than on the south side, with the lowest temperature found cospatial with the
  brightest line and X-ray regions \citep{blanton11}. Our results represent a
  significant new ingredient for modeling the interaction between the
  outflowing plasma and the external medium.

ALMA observations of central cluster galaxies have revealed a close
association between the X-ray cavities, \ha\ emission, and the molecular gas
\citep{mcnamara14,russell17,mcdonald12}. A large amount of cold molecular gas
($10^9-10^{10} M_\sun$) is common in BCG galaxies (e.g., Edge 2001;
Salome\&Combes 2003), but the origin of this gas is uncertain. A possible
explanation is that it is produced by the cooling of the hot plasma compressed
and uplifted by the expanding bubbles. Alternative scenarios exist, for example  the
lifting of dense gas out of the galaxy or the deposition of cold gas as a
result of a merger or a fly-by encounter.  The last possibility seems
unlikely because it requires an almost radial orbit at low velocity.  The
information on the distribution of the cold molecular gas in 3C~317 would be
of great importance in this respect, but it is still currently not available.

\section{Summary and conclusion}

MUSE data have provided new and detailed insights into the physical
properties of the WIM of UGC~9799, the central galaxy in the Abell cluster
2052, with  full 3D coverage of the X-ray cavities. The ionized gas forms a
filamentary structure surrounding the northern radio lobe. No emission is seen
instead  on the southern side with an asymmetry between the sides higher
than a factor 75. 

The gas velocity field is indicative of an expanding nebula
and we found regions where the emission lines are split, marking the locations
where the two sides project onto each other. The maximum observed velocity  is
265 \kms. The expansion speed of the nebula derived from these observations is
a factor  of $\sim$ 2 lower than previous indirect estimates based on X-ray
data. However, our measurement represents the instantaneous velocity instead
of the value averaged across its whole lifetime. 
 
From optical emission lines we also measured gas density (up to 270
cm$^{-3}$), estimated its mass (on the order of one million solar masses)
across the cavity and kinetic energy (on the order of $10^{53}$ erg). The
emission line ratios reveal that the warm gas is photoionized by the AGN
only within the central $\sim 3$ kpc, while they are inconsistent with AGN or
ionization of young stars  at greater distances. Because the emission lines are
generally quite narrow, 50 -- 100 \kms, shock ionization appears unlikely,
favoring ionization from cosmic rays instead.

The kinematics of the WIM described above clearly indicates that the radio
source has a profound impact on the external medium, confirming that radio
mode feedback is operating in A2052. However, the WIM is just the tip of the
iceberg of the various ISM phases. The X-ray data indicate that $\sim 10^{10}
M_\odot$  of the tenuous hot gas have been displaced. Furthermore, it is
likely that UGC 9799 harbors a large amount of cold and molecular gas
($10^8-10^9 M_\odot$), as commonly observed in the galaxies at the center of
cluster. The interplay between all phases of the ISM and the relativistic
radio plasma is a very complex phenomenon. The MUSE data provide us with
detailed information on the ionized gas. When coupled with those that can be
derived in the other observing bands these observations can be used to proceed
toward a comprehensive description of the energy exchange between the AGN and
the surrounding gas, tackling the issues of the origin, acceleration, and
ultimate fate of the cold and warm gas clouds.

\begin{acknowledgements}
We thank the referee for his/her constructive report.
Based on observations made with ESO Telescopes at the La Silla Paranal
Observatory under program ID 097.B-0766(A). This research has made use of
data obtained from the Chandra Data Archive. The National Radio Astronomy
Observatory is a facility of the National Science Foundation operated under
cooperative agreement by Associated Universities, Inc. Funding for SDSS has
been provided by the Alfred P. Sloan Foundation, the Participating
Institutions, the National Science Foundation, and the U.S. Department of
Energy Office of Science. B.B. acknowledge financial contribution from the
agreement ASI-INAF I/037/12/0.
\end{acknowledgements}

\bibliographystyle{aa} 

\end{document}